\documentstyle[aps,prd,twocolumn,epsf,floats]{revtex}
\begin{document}
\twocolumn[
  \begin{flushright}
  {\rm UCSD/PTH/98-36} \\
  {\rm HLRZ1998/62}
  \end{flushright}
  \vspace{-1cm}
  \begin{center}
  {\large\bf\ignorespaces
   The glueball spectrum from an anisotropic lattice study}\\
   \bigskip
   Colin~J.~Morningstar \\
   {\small\it Dept.~of Physics, University of California at San Diego,
    La Jolla, California 92093-0319} \\ 
   \bigskip
   Mike Peardon \\
   {\small\it John von Neumann-Institut f\"ur Computing and DESY,
    Forschungszentrum J\"ulich, D-52425 J\"ulich, Germany}\\
   \medskip
   {\small\rm (January 7, 1999)} \\
   \bigskip
  \begin{minipage}{5.5 true in} \small\quad
 The spectrum of glueballs below 4 GeV in the SU(3) pure-gauge theory is
 investigated using Monte Carlo simulations of gluons on several
 anisotropic lattices with spatial grid separations ranging from 0.1
 to 0.4 fm.  Systematic errors from discretization and finite volume
 are studied, and the continuum spin quantum numbers are identified.
 Care is taken to distinguish single glueball states from two-glueball
 and torelon-pair states.  Our determination of the spectrum significantly
 improves upon previous Wilson action calculations.
 \bigskip

 PACS number(s): 12.38.Gc, 11.15.Ha, 12.39.Mk
\end{minipage}
\end{center}
\vspace{6mm}
]

\section{Introduction}
\label{sec:intro}

The gluon self-coupling in quantum chromodynamics (QCD) suggests the
existence of glueballs, bound states of mainly gluons.  Incontrovertible
experimental evidence for their existence remains elusive, however.  A primary
reason for this is the difficulty in extracting the properties of
glueballs from the QCD lagrangian.  Investigating glueball physics requires an 
intimate knowledge of the confining QCD vacuum, and such knowledge cannot be 
obtained using standard perturbative techniques.  Numerical simulations of the 
theory on a space-time lattice currently provide the most reliable means of
studying glueballs.  However, correlation functions of gluonic exitations are
notoriously difficult quantities to measure in Monte-Carlo simulations,
requiring large-scale computer resources when applying standard
stochastic techniques.  Recently, the use of spatially coarse, 
temporally fine lattices and improved actions was demonstrated
to dramatically increase the efficiency of glueball 
simulations\cite{effglue}.

The objective in this paper is to apply the techniques of Ref.~\cite{effglue}
to substantially improve our knowledge of the glueball spectrum 
in the pure SU(3) gauge theory.  Detailed information on this spectrum is
important for validating models of confined gluons and may help
focus experimental searches for candidate glueball resonances. 
We also view this calculation as a necessary first step before
attempting to include the effects of quarks.  
Note that unlike the quenched approximation for mesons and baryons, the pure 
glue theory is a physical quantum field theory with a unitary S-matrix.  
First, we perform six simulations for spatial lattice spacings ranging from
0.1 to 0.4 fm to determine the energies of the lowest-lying stationary
states in all of the symmetry channels allowed on a cubic lattice.  In
many channels, we also determine the energies of the first-excited
states.  Our goal is then to extract the masses of as many low-lying
glueballs as possible from the 141 measurements which were made.  Since the
spectrum of glue defined in a box with periodic boundary conditions
includes not only single glueball states, but also states consisting
of several glueballs and/or torelons (gluon excitations which wrap
around the toroidal lattice), a means of identifying the single
glueballs must be employed to prune away all of the other 
unwanted states.  An additional small-volume simulation is done
to assist in this identification and to study the systematic errors
from finite volume.  Finally, discretization errors are treated by
extrapolating the energies to the continuum limit and determining
the continuum spin quantum numbers.  The end result is a nearly
complete survey of the glueball spectrum in the pure gauge theory
below 4 GeV.  We find a total of thirteen glueballs; two other
tentative candidates are also located.  With the exception of the
light glueballs in the $0^{++}$ sector, our results significantly
improve upon those from previous studies\cite{ukqcd,teper,berg}
of the complete low-lying glueball spectrum.

This paper is organized as follows.  The details of the simulations,
including the construction of the glueball operators, the generation
of the gauge-field configurations, the extraction of 
energies from Monte Carlo estimates of the correlation functions,
and the lattice spacing determinations in terms of the hadronic
scale $r_0$, are described in Sec.~\ref{sec:simdetails}.  All of our
energy estimates in terms of the inverse temporal lattice spacing
are presented in this section.  In Sec.~\ref{sec:twoglue}, the 
differentiation of single glueball states from two-glueball and
torelon-pair states is discussed.  Systematic errors from finite volume
are studied in Sec.~\ref{sec:finvol}.  The removal of lattice spacing
errors, including the extrapolations to the continuum limit and the
identification of the continuum spin quantum numbers, is described
in Sec.~\ref{sec:lattspac}.  We also discuss the problematical scalar
states in this section and cite ongoing efforts to reduce their
discretization errors.  Sec.~\ref{sec:discuss} presents a discussion of
the spectrum, and our findings are summarized with an
outline of future work in the concluding Sec.~\ref{sec:conclude}.

\section{Simulation Details}
\label{sec:simdetails}

Our glueball mass determinations rely on numerical simulations of gluons
on a hypercubic Euclidean space-time lattice with spatial and temporal
spacings $a_s$ and $a_t$, respectively.   The gluons are described by the
improved action $S_{I\!I}$ used in Ref.~\cite{effglue}.  The couplings in the
action depend on two parameters, $\beta$ and $\xi$, and are determined using
a combination of (tree-level) perturbation theory and mean-field theory,
implemented by renormalizing the spatial link variables 
$U_j(x)\rightarrow U_j(x)/u_s$, where $u_s$ is given by the fourth root of the
average plaquette\cite{lm}.  The temporal link variables are not renormalized.
The lattice anisotropy $a_s/a_t$ is given by $\xi$ at tree-level in
perturbation theory.  This action is intended for use with $a_t\ll a_s$, has
$O(a_s^4,a_t^2,\alpha_s a_s^2)$ discretization errors, where $\alpha_s$
is the QCD coupling, and couples only nearest-neighbor time slices,
ensuring the free-gluon propagator has no spurious modes.  In all cases,
glueball effective masses are seen to converge monotonically from above. 
This is a very desirable feature since it
validates the use of variational techniques to minimize excited-state
contributions to the glueball correlation functions.  Such techniques are
crucial for precise glueball mass extractions.

On a simple cubic lattice, zero-momentum stationary glue states are
characterized by their transformation properties under the octahedral
point group $O$, combined with parity and charge conjugation operations. 
$O$ has 24 elements (which are all proper rotations) that
fall into five conjugacy classes; the single-valued irreducible 
representations are labeled $A_1$, $A_2$, $E$, $T_1$, and $T_2$, 
(Sch\"onfliess notation\cite{schoenflies}) and have dimensions
1, 1, 2, 3, and 3, respectively\cite{cornwell}.  The inclusion of parity
results in the symmetry group known as $O_h = O\otimes C_i$, where
$C_i$ denotes the two-element group consisting of the identity operation
and spatial inversion.  The conventional labels for the irreducible
representations of $O_h$ are obtained from those of $O$ by appending
a subscript $g$ for representations corresponding to states which are
even under parity and $u$ for odd parity representations.  However, we
shall use a slightly different notation.  Instead of the subscripts $g$
and $u$, we use superscripts $+$ and $-$, respectively, to indicate the
eigenvalue $P$ of parity.  Glueball states are also eigenstates of charge
conjugation.  We denote the eigenvalue of charge-conjugation parity by $C$,
as usual, and introduce an additional superscript in the representation
labels.  We refer to the full symmetry group of zero-momentum glueball
states on a simple cubic lattice as $O_h^C$ or $O^{PC}$; the irreducible
representations are labeled $A_1^{PC}$, $A_2^{PC}$, $E^{PC}$, $T_1^{PC}$,
and $T_2^{PC}$.  For convenience, we use $R$ when referring
to these labels in general.  Note that when we use one of these labels
to identify a particular state, we refer to the lowest-lying
zero-momentum state in the symmetry channel indicated by the representation
label.  The first-excited state in a particular symmetry channel
will be denoted by the representation label with an asterisk.

The mass of a glueball $G$ having spin $J$, parity $P$, and charge-conjugation
parity $C$ can be extracted from the large-$t$ behavior of a lattice-regulated
correlation function $C(t)=\langle 0\vert \bar{\Phi}^{(R)}(t)
\, \bar{\Phi}^{(R)}(0)\vert 0\rangle$, where $R$ is any irreducible
representation of $O_h^C$ occurring in the subduced representation
$J^{PC}\!\downarrow\! O_h^C$, and 
$\bar{\Phi}^{(R)}(t)=\Phi^{(R)}(t)-\langle 0\vert \Phi^{(R)}(t)\vert 0\rangle$ 
is a gauge-invariant, translationally-invariant, vacuum-subtracted, real
operator capable of creating a glueball from the QCD vacuum $\vert 0\rangle$.
As the temporal separation $t$ becomes large, this correlator tends to a single 
decaying exponential $\lim_{t\rightarrow \infty} C(t) = Z \exp(-m_G t)$, where 
$m_G$ is the energy of the lowest-lying state which can be created by the 
operator $\bar{\Phi}^{(R)}(t)$.  To determine $m_G$, the correlator $C(t)$ must 
be calculated for large enough $t$ such that it is well approximated by its 
asymptotic form.  Unfortunately, stochastic fluctuations in $C(t)$ remain 
roughly constant with $t$ while the signal falls rapidly and hence,
the use of a glueball operator for which $C(t)$ attains its asymptotic form
as quickly as possible is crucial for reliably extracting $m_G$.  The energies
of excited states in representation $R$ can be obtained from
the large-$t$ behavior of a matrix of correlation functions
$C_{ij}(t)=\langle 0\vert \bar{\Phi}_i^{(R)}(t)
\, \bar{\Phi}_j^{(R)}(0)\vert 0\rangle$, where each of the glueball
operators $\bar{\Phi}_i^{(R)}(t)$ transforms as $R$ under all symmetry
operations.  Again, it is very important to use operators for which the
matrix elements $C_{ij}(t)$ attain their expected asymptotic forms for $t$
as small as possible.

\begin{figure}[t]
\begin{center}
\leavevmode
\epsfxsize=\columnwidth\epsfbox[24 241 570 611]{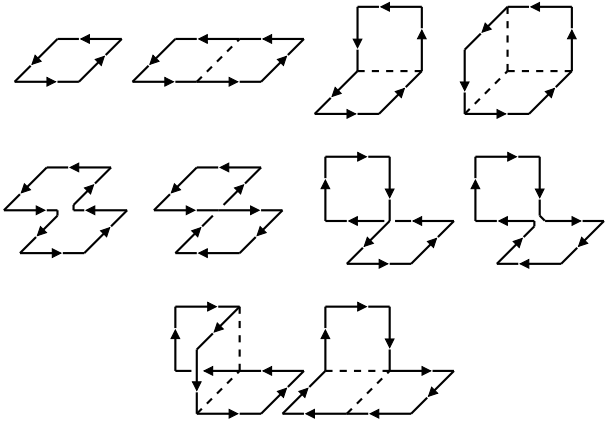}
\end{center}
\caption[figloops]{
  The Wilson loop shapes used in making the basic glueball operators.}
\label{fig:loops}
\end{figure}

Such operators can be constructed by exploiting link-smearing and variational
techniques as previously described in Ref.~\cite{effglue}.  For each
irreducible representation $R$, glueball operators on a given time-slice
are constructed in a sequence of three steps.  First, a set of six
smearing schemes are applied to the spatial link variables.  Each scheme is
a sequence of single-link and double-link mappings which depend on
parameters $\lambda_s$ and $\lambda_f$, respectively.  We use the same six
schemes described in Ref.~\cite{effglue}.  Secondly, a set of basic real
operators $\phi^{(R)}_{\alpha}(t)$ is constructed using linear combinations of 
gauge-invariant, path-ordered products of the smeared link matrices about
various closed spatial loops; each such linear combination is invariant under
spatial translations and transforms according to the irreducible representation
$R$. The loop-shapes employed in our calculation, shown in 
Fig.~\ref{fig:loops}, are chosen for their ease of computation; all orientations
of these operators can be computed very efficiently by first storing the
untraced products of links around the twelve spatial plaquettes stemming
from each site on a time-slice, then tracing the appropriate products of these
objects.  Both single and double windings around the paths are used; this
allows us to double the number of raw operators with only a small increase
in computational effort.
For each symmetry channel except $A_2^{-+}$, four irreducible combinations
are then chosen and applied to the smeared links from each of the six
schemes, yielding a total of 24 basic operators in each channel.  For
the $A_2^{-+}$, only the last shape in Fig.~\ref{fig:loops} can be used
and produces a total of 12 basic operators.
Lastly, the glueball operators $\Phi^{(R)}(t)$ are formed from
linear combinations of the basic operators, $\Phi^{(R)}(t)=\sum_\alpha
v_\alpha^{(R)} \phi^{(R)}_{\alpha}(t)$, where the coefficients 
$v^{(R)}_\alpha$ are determined using the variational method.  This involves
first obtaining Monte Carlo estimates of the large correlation matrix 
\begin{equation}
\tilde C_{\alpha\beta}(t) = \sum_\tau
            \langle 0\vert\bar{\phi}^{(R)}_{\alpha}(\tau\!+\!t) \:
            \bar{\phi}^{(R)}_{\beta}(\tau)\vert 0 \rangle,
\label{eq:gluecorr}
\end{equation}
where $\bar{\phi}^{(R)}_{\alpha}(t) = \phi^{(R)}_{\alpha}(t) - \langle 0 \vert
\phi^{(R)}_{\alpha}(t)\vert 0 \rangle$. In practise, this vacuum subtraction is 
only performed for the $A_1^{++}$ channel as the expectation value vanishes
identically in all other sectors. 
The coefficients $v^{(R)}_\alpha$ are 
then chosen to minimize the effective mass
\begin{equation}
\tilde m(t_D) = - \frac{1}{t_D}\ln\left[
\frac{\sum_{\alpha\beta}
 v^{(R)}_\alpha v^{(R)}_\beta\ \tilde C_{\alpha\beta}(t_D)}
{\sum_{\alpha\beta}
 v^{(R)}_\alpha v^{(R)}_\beta\ \tilde C_{\alpha\beta}(0)}
\right],
\end{equation}
where the time separation for optimization is fixed to $t_D=1$.
Other values of $t_D$ are used as consistency checks.
Let ${\bf v}^{(R)}$ denote a column vector whose elements are the optimal
values of the coefficients $v^{(R)}_\alpha$.  This vector satisfies
the eigenvalue equation
\begin{equation}
\tilde C(t_D)\ {\bf v}^{(R)} = e^{-t_D\tilde m(t_D)}\ \tilde C(0)
\ {\bf v}^{(R)}, \label{eq:Variation}
\end{equation}
and the eigenvector ${\bf v}_0^{(R)}$ corresponding to the lowest effective mass
$\tilde m_0(t_D)$ yields the coefficients $v^{(R)}_{0\alpha}$
for the operator $\Phi^{(R)}_0(t)$ which, under ordinary circumstances, best
overlaps the lowest-lying glueball $G_0$ in the channel of interest.  A sequence
of operators $\Phi^{(R)}_1(t), \Phi^{(R)}_2(t),\dots$ which predominantly
overlap excited glueball states can also be constructed using the higher-mass
eigenvectors of Eq.~(\ref{eq:Variation}).

\begin{table}[t]
\begin{center}
\caption[tabone]{
 The glueball simulation parameters.  Values for the coupling $\beta$,
 input aspect ratio parameter $\xi$, the mean-link parameter $u_s^4$,
 the single-link smearing weight $\lambda_s$, the two-link smearing
 weight $\lambda_f$, and the lattice sizes are listed.  Results
 for the hadronic scale $r_0$ in terms of the lattice spacing $a_s$
 are also given.  The approximate spatial lattice spacings $a_s$ are
 determined assuming $r_0^{-1}=410(20)$ MeV.
\label{tab:runparams}}
\begin{center}
\begin{tabular}{cccccrllc}
$\beta$ & $\xi$ & $u_s^4$ & $\lambda_s$ & $\lambda_f$
  & Lattice & \hspace{1ex} $r_0/a_s$ 
  & \hspace{1ex} $a_s/r_0$ & \hspace{-3ex} $a_s$ (fm)\\ \hline
1.7 &  5 & 0.295 & 0.1 & 0.5 & $6^3\times30$ & 1.224(1)& 0.8169(9) & 0.39\\
1.9 &  5 & 0.328 & 0.1 & 0.5 & $6^3\times30$ & 1.375(2)& 0.727(1)  & 0.35\\
2.2 &  5 & 0.378 & 0.1 & 0.5 & $8^3\times40$ & 1.761(2)& 0.5680(5) & 0.27\\
2.4 &  5 & 0.409 & 0.1 & 0.5 & $8^3\times40$ & 2.180(6)& 0.459(1)  & 0.22\\
2.5 &  5 & 0.424 & 0.1 & 0.5 & $10^3\times50$& 2.455(6)& 0.407(1)  & 0.20\\
3.0 &  3 & 0.500 & 0.4 & 0.5 & $15^3\times45$& 4.130(24)&0.2421(14)& 0.12
\end{tabular}
\end{center}
\end{center}
\end{table}

Monte Carlo estimates of the correlator matrix elements given in 
Eq.~(\ref{eq:gluecorr}) were obtained for all 20 irreducible representations
in five simulations using an input aspect ratio parameter
$\xi=5$.  The values for the coupling $\beta$, mean-link parameter
$u_s$, smearing weights $\lambda_s$ and $\lambda_f$, and the
lattice sizes used in these runs are listed in Table~\ref{tab:runparams}.
An additional run at a smaller lattice spacing ($\sim 0.12$ fm) and using
$\xi=3$ was done for the $A_1^{++}$, $E^{++}$, and $T_2^{++}$ representations
only.  A smaller-$a_s$ measurement helped to obtain a reliable
continuum-limit extrapolation for the troublesome $A_1^{++}$ state.
The input parameters for this run are also given in
Table~\ref{tab:runparams}.  All computations were carried out on
DEC Alpha and Sun Ultrasparc workstations.
Configuration ensembles were generated using
Cabibbo-Marinari (CM) pseudo-heatbath and SU(2) subgroup over-relaxation
(OR) methods.  Link variables were updated in serial order on the lattice.
Three compound sweeps were performed between measurements, where a compound
sweep is one CM updating sweep followed by $n_{\rm OR}$ OR sweeps.  The
measurements were averaged into bins of $n_{\rm mb}$, and $n_{\rm bins}$
bins were obtained.  For the $\beta=3.0$, $\xi=3$ run, $n_{\rm OR}=5$,
$n_{\rm mb}=50$, and $n_{\rm bins}=80$.  For the $\beta=2.5$, $\xi=5$ run,
$n_{\rm OR}=5$, $n_{\rm mb}=20$, and $n_{\rm bins}=318$.  For all of the
other simulations, $n_{\rm OR}=3$, $n_{\rm mb}=100$, and $n_{\rm bins}=100$.
Crude checks for residual autocorrelations were done
by over-binning by factors of two and four; in all cases,
statistical error estimates remained unchanged.

\begin{table*}[t]
\newcommand{\nomeas}{\hspace{1.6ex}-----}
\begin{center}
\begin{minipage}{4.5 true in}
\caption[tabtwo]{
 Glue energy estimates in terms of $a_t^{-1}$ for the five
 $\xi=5$ simulations.  The levels are labeled by the irreducible
 representations of the cubic point group under which their corresponding
 stationary states transform.  First-excited states in a representation
 are indicated by an asterisk.
\label{tab:masses}}
\begin{center}
\begin{tabular}{llllll}
      & $\beta=1.7$ & $\beta=1.9$ & $\beta=2.2$
      & $\beta=2.4$ & $\beta=2.5$  \\
\hline
 $A_1^{++}$      & 0.578(5) & 0.475(4) & 0.362(3) & 0.303(3) & 0.288(4) \\
 $A_1^{\ast ++}$ & 1.19(2)  & 0.92(3)  & 0.697(6) & 0.569(4) & 0.511(7) \\
 $A_2^{++}$      & 1.43(2)  & 1.27(3)  & 1.018(12)& 0.824(8) & 0.713(14)\\
 $E^{++}$        & 0.924(8) & 0.844(6) & 0.667(4) & 0.538(3) & 0.472(4) \\
 $E^{\ast ++}$   & 1.29(2)  & 1.093(12)& 0.878(7) & 0.723(9) & 0.652(8) \\
 $T_1^{++}$      & 1.55(2)  & 1.32(2)  & 1.00(2)  & 0.834(4) & 0.728(8) \\
 $T_1^{\ast ++}$ & 1.73(3)  & 1.52(2)  & 1.23(4)  & 0.909(15)& 0.823(13)\\
 $T_2^{++}$      & 1.103(8) & 0.918(7) & 0.686(4) & 0.542(2) & 0.477(3) \\
 $T_2^{\ast ++}$ & 1.41(2)  & 1.228(12)& 0.938(5) & 0.730(8) & 0.660(6) \\[3mm]
 $A_1^{-+}$      & 1.31(2)  & 1.06(5)  & 0.756(14)& 0.605(11)& 0.522(7) \\
 $A_1^{\ast -+}$ & 1.86(9)  & 1.47(4)  & 1.08(2)  & 0.836(9) & 0.72(2)  \\
 $A_2^{-+}$      & \nomeas  & 1.63(5)  & 1.29(2)  & 1.036(12)& 0.93(1)  \\
 $E^{-+}$        & 1.38(2)  & 1.167(14)& 0.874(5) & 0.698(4) & 0.611(7) \\
 $E^{\ast -+}$   & 1.85(5)  & 1.49(3)  & 1.085(9) & 0.890(6) & 0.782(13)\\
 $T_1^{-+}$      & 1.56(2)  & 1.42(2)  & 1.155(9) & 0.873(13)& 0.82(1)  \\
 $T_2^{-+}$      & 1.297(14)& 1.148(10)& 0.882(5) & 0.695(4) & 0.619(3) \\
 $T_2^{\ast -+}$ & 1.60(2)  & 1.39(2)  & 1.087(9) & 0.880(5) & 0.771(9) \\[3mm]
 $A_1^{+-}$      & \nomeas  & 1.67(5)  & 1.32(2)  & 1.062(14)& 0.94(1)  \\
 $A_2^{+-}$      & \nomeas  & 1.29(2)  & 0.999(11)& 0.796(7) & 0.700(14)\\
 $E^{+-}$        & \nomeas  & 1.52(3)  & 1.207(37)& 0.929(8) & 0.82(1)  \\
 $T_1^{+-}$      & 1.186(13)& 1.053(8) & 0.819(4) & 0.652(5) & 0.590(5) \\
 $T_1^{\ast +-}$ & 1.55(3)  & 1.297(14)& 1.025(8) & 0.794(9) & 0.73(2)  \\
 $T_2^{+-}$      & \nomeas  & 1.330(16)& 0.983(17)& 0.801(4) & 0.701(7) \\
 $T_2^{\ast +-}$ & \nomeas  & 1.45(2)  & 1.10(3)  & 0.929(15)& 0.83(1)  \\[3mm]
 $A_1^{--}$      & 1.83(7)  & 1.61(5)  & 1.354(20)& 1.04(3)  & 0.98(1)  \\
 $A_2^{--}$      & \nomeas  & 1.65(6)  & 1.201(18)& 0.96(3)  & 0.82(2)  \\
 $E^{--}$        & 1.59(3)  & 1.38(2)  & 1.094(11)& 0.875(6) & 0.78(1)  \\
 $T_1^{--}$      & 1.72(4)  & 1.46(2)  & 1.07(3)  & 0.877(5) & 0.760(9) \\
 $T_2^{--}$      & 1.63(3)  & 1.41(2)  & 1.114(8) & 0.886(6) & 0.77(1)
\end{tabular}
\end{center}
\end{minipage}
\end{center}
\end{table*}

In the final analysis phase, the glue energies $m_G$ were extracted 
using a two-step procedure.  First, the large correlation
matrices in each channel were reduced to smaller 
$3\times 3$ matrices $C_{AB}(t)$ for $A,B = 0,1,2$ using the variational
coefficients of the three lowest mass eigenstates 
of Eq.~(\ref{eq:Variation}):
\begin{equation}
C_{AB}(t)=\sum_\tau \langle 0\vert \bar{\Phi}_A^{(R)}(\tau\!+\!t)
\ \bar{\Phi}^{(R)}_B(\tau)\vert 0\rangle.
\end{equation}
Secondly, the expected large-$t$ functional forms were fit to the
elements of these optimized correlators.  To obtain an estimate of the energy
$m_{G_0}$ (in terms of $a_t^{-1}$) of the lowest-lying state in each channel, 
a single exponential was fit to the ground-state correlator $C_{00}(t)$
for $t=t_{\rm min},\dots,t_{\rm max}$:
\begin{equation}
  C_{00}(t) =  Z_{00}
        \left\{ e^{ -m_{G_0}t} + e^{-m_{G_0}(T-t)} \right\},
\label{eq:GGcorrelator}
\end{equation}
where $T$ was the temporal extent of the periodic lattice.  To determine the
energies $m_{G_p}$ of the excited states and another estimate of $m_{G_0}$,
the $M\times M$ optimized correlator matrix was also fit for 
$t=t_{\rm min},\dots,t_{\rm max}$ using the form
\begin{equation}
  C_{AB}(t) = \sum_{p=0}^{M-1}   Z_{Ap} Z_{Bp}
        \left\{ e^{ -m_{G_p}t} + e^{-m_{G_p}(T-t)} \right\},
\label{eq:MbyMcorrelator}
\end{equation}
for $M=2,3$.  Various fit regions $t_{\rm min}$ to $t_{\rm max}$ were used
to check for consistency in the extracted values for the masses.
Best-fit values were obtained using the correlated $\chi^2$ method.
Error estimates were calculated using a $1024-$point bootstrap procedure;
in all cases, error estimates were nearly symmetric about
the central best-fit values and were thus averaged to simplify presentation.
Our fits are far too numerous to list here; additional details are available
from the authors upon request.

Our final estimates of the glue energies in terms of $a_t^{-1}$ for
the $\xi=5$ simulations are presented in Table~\ref{tab:masses}; energy
estimates from the $\xi=3$ run are listed in Table~\ref{tab:massesb}.
In order to convert these
results into physical units, the lattice spacings $a_t$ must be determined for
each simulation.  The hadronic scale parameter $r_0$ \cite{sommerscale} 
defined in terms of the force between static quarks by 
$[r^2 dV(\vec{r})/dr]_{r=r_0}=1.65$, where $V(\vec{r})$ is the static-quark 
potential, is a useful quantity for this purpose. The values for $r_0$ in terms 
of $a_s$ corresponding to each glueball simulation were determined by measuring
the static-quark potential in separate simulations. The results are listed in
Table~\ref{tab:runparams}.  Further details concerning the calculation
of $r_0/a_s$ are given in Ref.~\cite{effglue}.  Note that in computing
$r_0/a_s$, the input value $\xi$ was used for the aspect ratio $a_s/a_t$.
The consequences of doing so are discussed in Sec.~\ref{sec:contlimit}.

\begin{table}[t]
\caption[tabthree]{
 Glue energy estimates in units of $a_t^{-1}$ from the
 $\beta=3.0$, $\xi=3$ simulation.
\label{tab:massesb}}
\begin{center}
\begin{minipage}{1.7 true in}
\begin{tabular}{ll}
 $A_1^{++}$   & 0.318(4) \\
 $E^{++}$     & 0.476(3) \\
 $T_2^{++}$   & 0.476(3)
\end{tabular}
\end{minipage}
\end{center}
\end{table}

\section{Glueball identification}
\label{sec:twoglue}

The spectrum in a box with periodic boundary conditions includes not only 
single glueball states, but also states consisting of several glueballs
and/or torelons (gluon excitations which wrap around the toroidal
lattice).  We expect that the operators used in our correlators
will couple most strongly to the single glueball states, but we cannot
be certain that mixings with the multi-glueball and torelon states will
be negligible.  Recall that the asymptotic behavior of the correlation
function associated with an operator $\Phi(t)$ is dominated by the
lowest-lying eigenstate which mixes with $\Phi$.  If a multi-glueball
or torelon state has a lower energy than the lightest glueball
in a given symmetry channel, then the possibility exists that the
energy we extract from the asymptotic decay of the correlator will be
that of the multi-glueball or torelon state.  Thus, a means of 
differentiating the single glueball states from all other states
is required.

First, given mass estimates of the lowest-lying few glueballs, one can
easily determine the approximate locations in a given symmetry channel
of the two-glueball states having zero total momentum.  If the simulation
results in that channel lie significantly below the two-glueball energy
estimates, one can almost certainly rule out a multi-glueball interpretation.

Secondly, one can study the manner in which each energy level
changes as the lattice volume is varied.  The energy of a single glueball
state depends on the lattice volume in a markedly different way from
that of a multi-glueball or torelon state.  

A third possibility is to
include additional operators in the correlation matrices which are
expected to couple much more strongly with the two-glueball and torelon
states.  The construction of operators which best overlap the
lowest-lying eigenstates of interest using the variational method then
involves not only single glueball operators, but also the new two-glueball
and torelon operators.  The coefficients obtained from the variational
optimization can be used to estimate the mixings of the additional
operators with the low-lying eigenstates of interest.  
If the mixings of the additional two-glueball and torelon
operators with an eigenstate are very small, a single glueball interpretation
is assured; in such a case, the addition of the new operators does not
affect the extracted energy.  If the mixings are significant, the addition
of the two-glueball/torelon operators will lower the extracted energy,
ruling out a single glueball interpretation. 

Ideally, it would be best to apply all three of these methods.  However,
for this initial scan of the glueball spectrum, we decided for reasons of
simplicity to rely mainly on the first method.  Having obtained the
lowest-lying one or two states in each symmetry channel, we determined
the approximate locations of the two-glueball and torelon states.
Simulation results lying near or above these thresholds were then excluded
from further consideration. In other words, we used the
two-glueball and torelon thresholds as filters to remove possibly extraneous
states.  One additional simulation was done to study systematic errors
from finite volume.  This run also served to confirm the single glueball
nature of the states lying well below the two-glueball/torelon thresholds.
Note that the levels lying near or above these thresholds cannot be ruled
out as single glueball states; rather, one can say only that the
interpretation of such states requires additional information.

\subsection{Two-glueball states}

In order to identify the genuine
single glueball states, we first determined the approximate locations of the
two-particle states using the mass estimates of the lightest few glueballs.
In estimating these locations, we assumed that the energy of a two-glueball
state was given by 
\begin{equation}
 E_{2G} \approx \sqrt{\vec{p}^2+m_1^2}+\sqrt{\vec{p}^2+m_2^2},
\label{eq:twoglueballmass}
\end{equation}
where $m_1$ is the rest mass of the glueball having momentum $\vec{p}$
and $m_2$ is the rest mass of the other glueball which has momentum $-\vec{p}$.
Note that on a periodic lattice having $N_s$ sites in each of the three
spatial directions, the allowed momenta are discrete
$\vec{p}=2\pi(n_x,n_y,n_z)/L$, where $L=a_sN_s$ and $n_x$, $n_y$, and
$n_z$ are integers satisfying $-N_s/2<n_j\leq N_s/2$.  
The above energy estimates neglect all interactions between the
two glueballs; this should not introduce serious error since we
expect these interactions to be short-ranged, being mediated by
scalar glueball exchange at large distances.
Eq.~(\ref{eq:twoglueballmass}) also assumes that the rest masses and
dispersion relations of the propagating glueballs are unaffected by
finite volume and finite lattice spacing errors.  We have verified this
assumption in the case of the scalar glueball for $n_x^2+n_y^2+n_z^2<9$
on an $8^3\times 40$ lattice at $\beta=2.4$ and $\xi=5$.

\begin{table}[t]
\caption[tabfour]{
 The little groups of $\vec{p}=2\pi(n_x,n_y,n_z)/L$ on a
 simple cubic lattice with periodic boundary conditions.  Note that
 $l,m,n \neq 0$ and $l\neq m$, $m\neq n$, and $l\neq n$. 
\label{tab:littlegroups}}
\begin{center}
\begin{minipage}{2.0 true in}
\begin{tabular}{cc}
 $(n_x,n_y,n_z)$ &  Little group \\
 \hline
   $(0, 0, 0)$   &    $O_h$      \\
   $(0, 0, n)$   &   $C_{4v}$    \\
   $(0, n, n)$   &   $C_{2v}$    \\
   $(n, n, n)$   &   $C_{3v}$    \\
   $(0, m, n)$   &    $C_s$      \\
   $(m, m, n)$   &    $C_s$      \\
   $(l, m, n)$   &    $C_1$
\end{tabular}
\end{minipage}
\end{center}
\end{table}

To facilitate our discussion of the two-glueball states, we first
point out some features of single, propagating glueballs in a finite box with
periodic boundary conditions. Here, glueball states are characterized by their 
transformation properties under $O_h^{1C}$, the simple cubic crystallographic 
space group $O_h^1$ extended to include charge-conjugation. The group $O_h^1$ 
is isomorphic to the semi-direct product of the group of pure (discrete) 
translations and the group of pure (discrete) rotations and reflections about a 
given center. Thus, a propagating glueball state may be classified by its
momentum $\vec{p}$ and by its transformation properties under the subgroup
of $O_h^{1C}$ which leaves $\vec{p}$ invariant (the little group of $\vec{p}$).
The little groups corresponding to various momentum orientations are listed
in Table~\ref{tab:littlegroups}.  Hence, the irreducible representations
of the little group may be used to identify propagating glueball states.
The little group varies with the momentum orientation.  This means that
the partitioning of the physical glueball states into the irreducible
representations of the lattice symmetry groups differs depending on the
momentum orientation.  For example, consider the $2^{++}$ glueball.
When at rest, three of the five polarizations of this glueball appear in
the $T_2^{++}$ representation of $O_h^C$, the little group of $\vec{p}=(0,0,0)$,
and two of its polarizations occur in the $E^{++}$ representation.
When $\vec{p}=(0,0,p)$ for $p\neq 0$, the five polarizations of the $2^{++}$
glueball (which are no longer eigenstates of parity) split across the $A_1^+$,
$B_1^+$, $B_2^+$, and $E^+$ representations of the little group $C_{4v}^C$.

We determined the lowest-lying two-glueball states in each symmetry
channel by repeating the following sequence of steps for each allowed
momentum vector $\vec{p}$ and each choice of two glueballs $G_1$ and $G_2$.
For the moment, assume that $G_1$ and $G_2$ are distinguishable.  Note that
$G_1$ and $G_2$ refer to the irreducible representations of $O_h^C$, the
little group for zero momentum.  First, we identified the little group
$L(\vec{p})$ of $\vec{p}$.  Secondly, the characters $\chi^{(G_1)}$ and
$\chi^{(G_2)}$ of the representations $G_1$ and $G_2$ were subduced into the
little group, yielding the characters $\chi^{(G_j)}\!\downarrow\!L(\vec{p})$,
which were then decomposed into the irreducible representations $\mu$ of the
little group: $\chi^{(G_j)}\!\downarrow\! L(\vec{p})=\sum_\mu c^{(G_j)}_\mu
\chi^{(\mu)}$.  Next, for each $\mu_1$ such that $c^{(G_1)}_{\mu_1}\neq 0$
and each $\mu_2$ such that $c^{(G_2)}_{\mu_2}\neq 0$, we formed the direct
product $\chi^{(\mu_1\otimes\mu_2)}$ to obtain the character
corresponding to the two-glueball state.  Since the total
momentum of this two-glueball state is zero, it can be characterized
by the irreducible representations of $O_h^C$.  A representation of $O_h^{C}$ 
was formed by constructing a set of coset representatives and applying
the method of induction, and the induced character
$\chi^{(\mu_1\otimes\mu_2)}\!\!\uparrow\!\!O_h^C$ was finally decomposed into
the irreducible representations of $O_h^C$.  When $G_1$ and $G_2$ were
indistinguishable, the above procedure was modified to include
Bose symmetrization.

\begin{figure}[t]
\begin{center}
\leavevmode
\epsfxsize=\columnwidth\epsfbox[26 87 497 651]{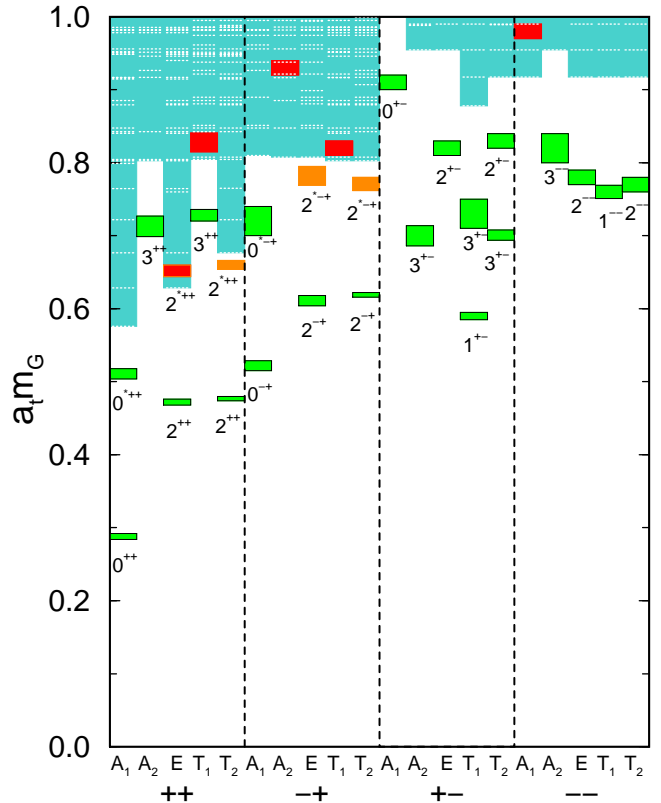}
\end{center}
\caption[figspinanal]{
  Comparison of the pure-glue spectrum obtained from the $\beta=2.5$, $\xi=5$
  simulation to the approximate locations of the two glueball states.
  The boxes are the simulation results; the standard deviations in these
  mass estimates are indicated by the vertical heights of the boxes.  The 
  dotted line segments shown in the upper shaded region indicate the approximate
  locations of states consisting of two free glueballs having zero total
  momentum.  All energies are in terms of $a_t^{-1}$.  The representations
  of the cubic point group which label the states are indicated along the
  horizontal axis.  The most likely $J^{PC}$ interpretations of the states
  are also shown.}
\label{fig:spinanal}
\end{figure}

This procedure was carried out using the rest energies obtained in the
$\beta=2.5$, $\xi=5$ simulation.  The lowest-lying two-glueball states
in each symmetry channel are listed in Table~\ref{tab:twoglueball}.
These levels, along with all higher lying levels, are shown as dashed line
segments in the shaded region in Fig.~\ref{fig:spinanal}.  Any energy
lying well below the shaded region in this figure can be safely
interpreted as a single glueball state (these are indicated by the
black-outlined green-filled boxes).  States lying slightly below (the
orange boxes with no outlines) or above (the red boxes) the two-glueball
thresholds must be regarded with
caution.  Again, we remind the reader that we cannot rule out a single
glueball interpretation for these levels; additional information is
needed to determine the nature of these states.  Since it is not our
intent in this paper to obtain such information, we exclude these levels
from the spectrum of single glueball states for the time being.

\begin{table}[t]
\caption[tabfive]{
 The lowest lying state consisting of two free glueballs in each
 symmetry channel for the $\beta=2.5$, $\xi=5$ simulation.  Each state is
 comprised of a glueball having a momentum $\vec{p}=2\pi(n_x,n_y,n_z)/L$,
 where $L=10a_s$, and another glueball having momentum $-\vec{p}$. 
\label{tab:twoglueball}}
\begin{center}
\begin{minipage}{3.0 true in}
\begin{tabular}{ccc}
 Channel  & Glueballs & $(n_x,n_y,n_z)$ \\
\hline
 $A_1^{++}$ & $(A_1^{++}, A_1^{++})$ & (0, 0, 0) \\
 $A_2^{++}$ & $(A_1^{++}, E^{++}  )$ & (0, 0, 1) \\
 $E^{++}$   & $(A_1^{++}, A_1^{++})$ & (0, 0, 1) \\
 $T_1^{++}$ & $(A_1^{++}, A_1^{++})$ & (0, 1, 2) \\
 $T_2^{++}$ & $(A_1^{++}, A_1^{++})$ & (0, 1, 1) \\
 $A_1^{-+}$ & $(A_1^{-+}, A_1^{++})$ & (0, 0, 0) \\
 $A_2^{-+}$ & $(A_1^{++}, T_2^{++})$ & (0, 0, 1) \\
 $E^{-+}$   & $(A_1^{++}, T_2^{++})$ & (0, 0, 1) \\
 $T_1^{-+}$ & $(A_1^{++}, E^{++}  )$ & (0, 0, 1) \\
 $T_2^{-+}$ & $(A_1^{++}, E^{++}  )$ & (0, 0, 1) \\
 $A_1^{+-}$ & $(A_1^{++}, T_1^{+-})$ & (0, 1, 2) \\
 $A_2^{+-}$ & $(A_1^{++}, T_1^{+-})$ & (0, 1, 1) \\
 $E^{+-}$   & $(A_1^{++}, T_1^{+-})$ & (0, 1, 1) \\
 $T_1^{+-}$ & $(A_1^{++}, T_1^{+-})$ & (0, 0, 0) \\
 $T_2^{+-}$ & $(A_1^{++}, T_1^{+-})$ & (0, 0, 1) \\
 $A_1^{--}$ & $(A_1^{++}, T_1^{+-})$ & (0, 0, 1) \\
 $A_2^{--}$ & $(A_1^{++}, T_1^{+-})$ & (0, 1, 1) \\
 $E^{--}$   & $(A_1^{++}, T_1^{+-})$ & (0, 0, 1) \\
 $T_1^{--}$ & $(A_1^{++}, T_1^{+-})$ & (0, 0, 1) \\
 $T_2^{--}$ & $(A_1^{++}, T_1^{+-})$ & (0, 0, 1)
\end{tabular}
\end{minipage}
\end{center}
\end{table}

\subsection{Torelon pairs}

Torelons are gluonic excitations which wind around the periodic boundaries
of the lattice.  They may be classified according to their behavior
under a set of discrete $Z_3$ symmetries of the SU(3) pure gauge theory.
The gauge action is invariant under multiplication of every link in
the $\mu$ direction ($\mu=0,1,2,3)$ at a given $\mu$ coordinate by the
same member of $Z_3$, the center of SU(3).  The torelon is an eigenstate
of the transfer matrix which transforms nontrivially under such
symmetry operations.  For example, the spectrum of glue on a periodic
lattice contains a torelon eigenstate which picks up a phase 
$\exp(2\pi i/3)$ and a state which picks up a phase $\exp(-2\pi i/3)$ under
the multiplication of every link in the $x$-direction at a given $x$
coordinate by the center member $\exp(2\pi i/3)$.  In fact,
there are three such pairs of modes corresponding to the $x$, $y$, and $z$
directions.  These torelons can also have a center-of-mass momentum
in the two spatial directions transverse to their flux direction.
For large $L$, a torelon at rest has an energy given approximately by
$\sigma L$, where $\sigma$ is the string tension from the static-quark
potential and $L$ is the spatial extent of the lattice.  We have
confirmed this in a simulation on an $8^3\times 24$ lattice at
$\beta=2.4$, $\xi=3$.   Hence, the torelon mass is strongly
dependent on the volume of the lattice.

Our glueball operators, being closed Wilson loops which do not wrap
around the lattice, are invariant under these center symmetry 
transformations.  This means that the glueball operators cannot create
a single torelon state, but the creation of torelon pairs of opposite
center charge is possible.  If the two torelons of opposite charge
do not substantially interact, then the lowest energy of such a state
is $2\sigma L$, which has the value 0.9 when placed on 
Fig.~\ref{fig:spinanal}.  This lies sufficiently high to discard
from consideration, even for the $A_1^{+-}$ state since a state of
two torelons of opposite center charge and total zero momentum must
be symmetric under charge conjugation.  Fortunately, a torelon pair
can be easily detected in a finite-volume study since their
energy depends strongly on $L$. An additional
simulation was done to measure the changes in all energy levels as
the lattice volume was reduced.  The results of this simulation are
presented in the next section.  No energy reductions of sufficient
magnitude were found to suggest that any of our states could be
interpreted as a torelon pair.

\section{Finite volume errors}
\label{sec:finvol}

\begin{figure}[t]
\begin{center}
\leavevmode
\epsfxsize=\columnwidth\epsfbox[68 154 562 668]{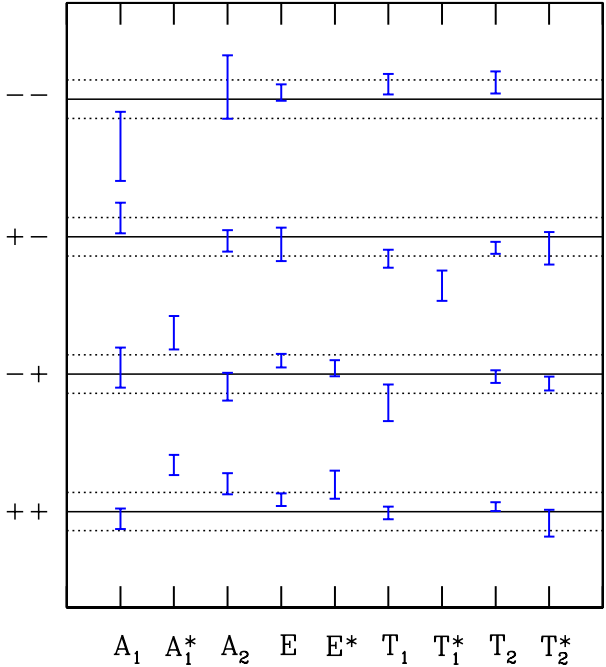}
\end{center}
\caption[figfseone]{Finite-volume effects on the results of the
  $\beta=2.4$, $\xi=5$ simulation.  Each point shows the fractional change 
  $\delta_G=1-m_G^S/m_G$ in the energy of a stationary state $G$,
  where $m_G$ is the energy of $G$ as measured on an $8^3\times 40$ lattice,
  and $m_G^S$ is the energy of $G$ as measured on a smaller
  $6^3\times 40$ lattice.  The state $G$ corresponding to a given point
  is specified by combining the representation label below the point
  along the horizontal axis with the $PC$ value shown to its left along
  the vertical axis.  The solid lines indicate $\delta_G=0$, the dotted
  lines above the solid lines indicate $\delta_G=0.02$, and the dotted
  lines below the solid lines indicate $\delta_G=-0.02$.}
\label{fig:fse}
\end{figure}

An additional simulation was done to measure the systematic errors
in our results from finite volume:  a run at $\beta=2.4$, $\xi=5$ on a
$6^3\times 40$ lattice.  The original $8^3\times 40$ lattice has a
spatial volume of (1.76 fm)$^3$, assuming that $a_s\sim 0.22$ fm
from $r_0^{-1}=410(20)$ MeV.  The additional simulation measures the changes
in the glueball masses as the volume is reduced from (1.76 fm)$^3$ to
(1.32 fm)$^3$.  The input parameters used in this small-volume run were
the same as those used in the larger volume simulation.

Let $m_G$ denote the energy of a state $G$ on the original
$8^3\times 40$ lattice, and $m_G^S$ denote the energy of $G$ as measured
on the smaller $6^3\times 40$ lattice.  The fractional change in the energy
is defined by $\delta_G=1-m_G^S/m_G$.  The results for these fractional
changes are shown in Fig.~\ref{fig:fse}.  Each point shows the fractional
change in the energy of a state $G$ specified by combining the 
representation label below the point on the horizontal axis with
the $PC$ value shown to its left on the vertical axis.
The solid lines indicate $\delta_G=0$, the dotted lines above the solid
lines indicate $\delta_G=0.02$, and the dotted lines below the solid lines
indicate $\delta_G=-0.02$.  The largest effects from finite volume
occur in the $A_1^{\ast ++}$ and $T_1^{\ast +-}$ states.  All other
changes are statistically consistent with zero, indicating that systematic
errors in these results from finite volume are negligible.  These results
confirm the single glueball nature of all of the states, with the
exception of the $A_1^{\ast ++}$, lying well below the two-glueball
thresholds (the black-outlined green-filled boxes in 
Fig.~\ref{fig:spinanal}).  Although the
change in the energy of the $A_1^{\ast ++}$ is not very large, it is
sufficient to warrant further study of this level.  For this reason,
we withhold judgment on whether or not this level is a single glueball.
Note that most of the states lying near or above the two-glueball thresholds
(the orange and red boxes in Fig.~\ref{fig:spinanal}) show very little
finite volume dependence, suggesting that these states might actually be
long-lived glueball resonances.  Further study would be required to resolve
this issue.

\section{Lattice spacing errors}
\label{sec:lattspac}

There are two aspects to the removal of systematic errors from finite
lattice spacing: extrapolation of the results to the continuum limit
$a_s\!\rightarrow\!0$, and the identification of the continuum spin
quantum numbers.  In this section, we first carry out the 
$a_s\!\rightarrow\!0$ extrapolations of the candidate single-glueball levels 
remaining after the analysis of Sec.~\ref{sec:twoglue}, then deduce their 
continuum spin content.

\subsection{Continuum limit extrapolations}
\label{sec:contlimit}

\begin{figure}[t]
\begin{center}
\leavevmode
\epsfxsize=\columnwidth\epsfbox[9 67 529 631]{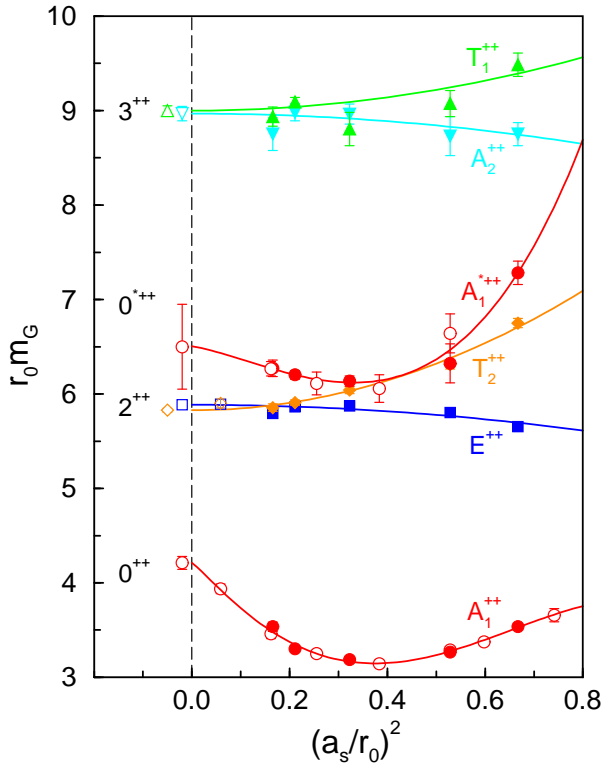}
\end{center}
\caption[figgluePP]{
  Mass estimates of the $PC=++$ glueballs in terms of $r_0$
  against the lattice spacing $(a_s/r_0)^2$.  The solid symbols
  indicate results from the $\xi=5$ simulations, and the open
  symbols on the right-hand side of the vertical dashed line indicate
  results from the $\xi=3$ simulations.  The solid curves are best fits
  to the simulation results for each state using $\varphi_1(a_s)$
  from Eq.~(\protect\ref{phiscalar}) for the $A_1^{++}$ and $A_1^{\ast ++}$
  levels and $\varphi_0(a_s)$ from Eq.~(\protect\ref{phifour}) for all
  other levels.  The open symbols on the left-hand
  side of the vertical dashed line show the extrapolations to the
  continuum limit using the best-fit forms.}
\label{fig:gluePP}
\end{figure}

\begin{figure}[t]
\begin{center}
\leavevmode
\epsfxsize=\columnwidth\epsfbox[9 67 529 631]{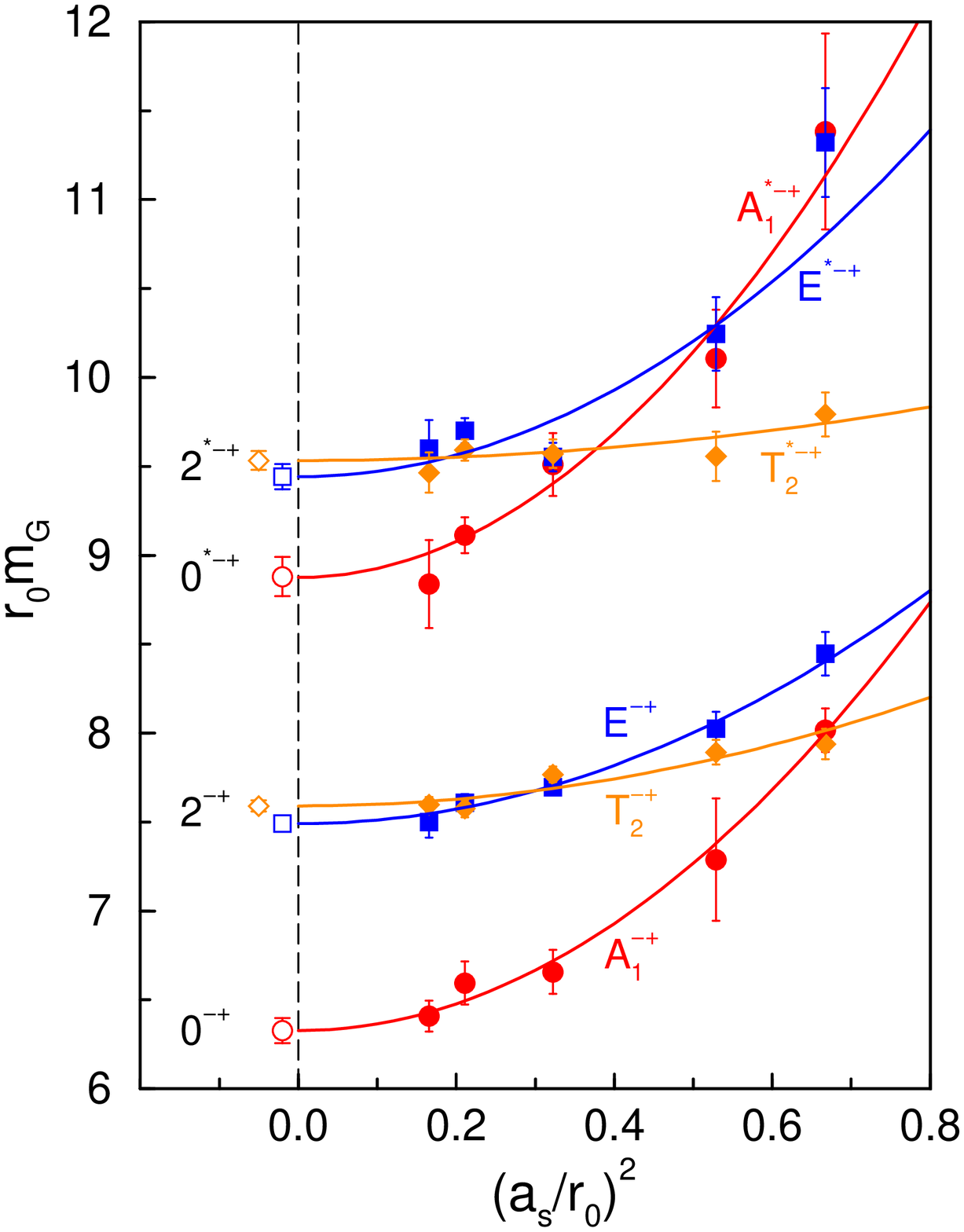}
\end{center}
\caption[figglueMP]{
  Mass estimates (solid symbols) of the $PC=-+$ glueballs in terms of $r_0$
  against the lattice spacing $(a_s/r_0)^2$.  The solid curves are best
  fits of $\varphi_0(a_s)$ from Eq.~(\protect\ref{phifour}) to the
  results for each state; the open symbols are the continuum limit
  extrapolations.}
\label{fig:glueMP}
\end{figure}

\begin{figure}[t]
\begin{center}
\leavevmode
\epsfxsize=\columnwidth\epsfbox[9 67 529 631]{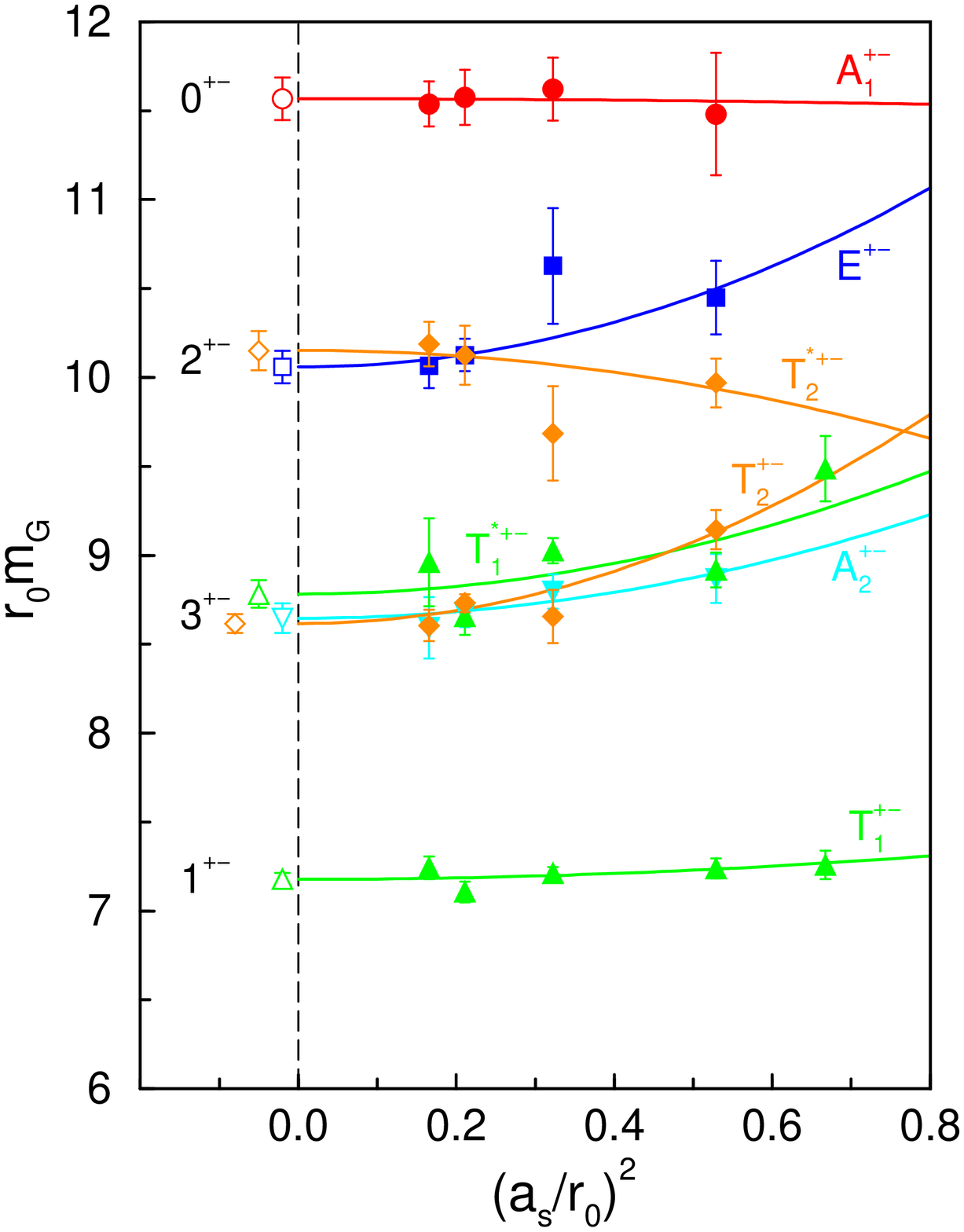}
\end{center}
\caption[figgluePM]{
  Mass estimates (solid symbols) of the $PC=+-$ glueballs in terms of $r_0$
  against the lattice spacing $(a_s/r_0)^2$.  The solid curves are best
  fits of $\varphi_0(a_s)$ from Eq.~(\protect\ref{phifour}) to the
  results for each state; the open symbols are the continuum limit
  extrapolations.}
\label{fig:gluePM}
\end{figure}

\begin{figure}[t]
\begin{center}
\leavevmode
\epsfxsize=\columnwidth\epsfbox[9 67 529 631]{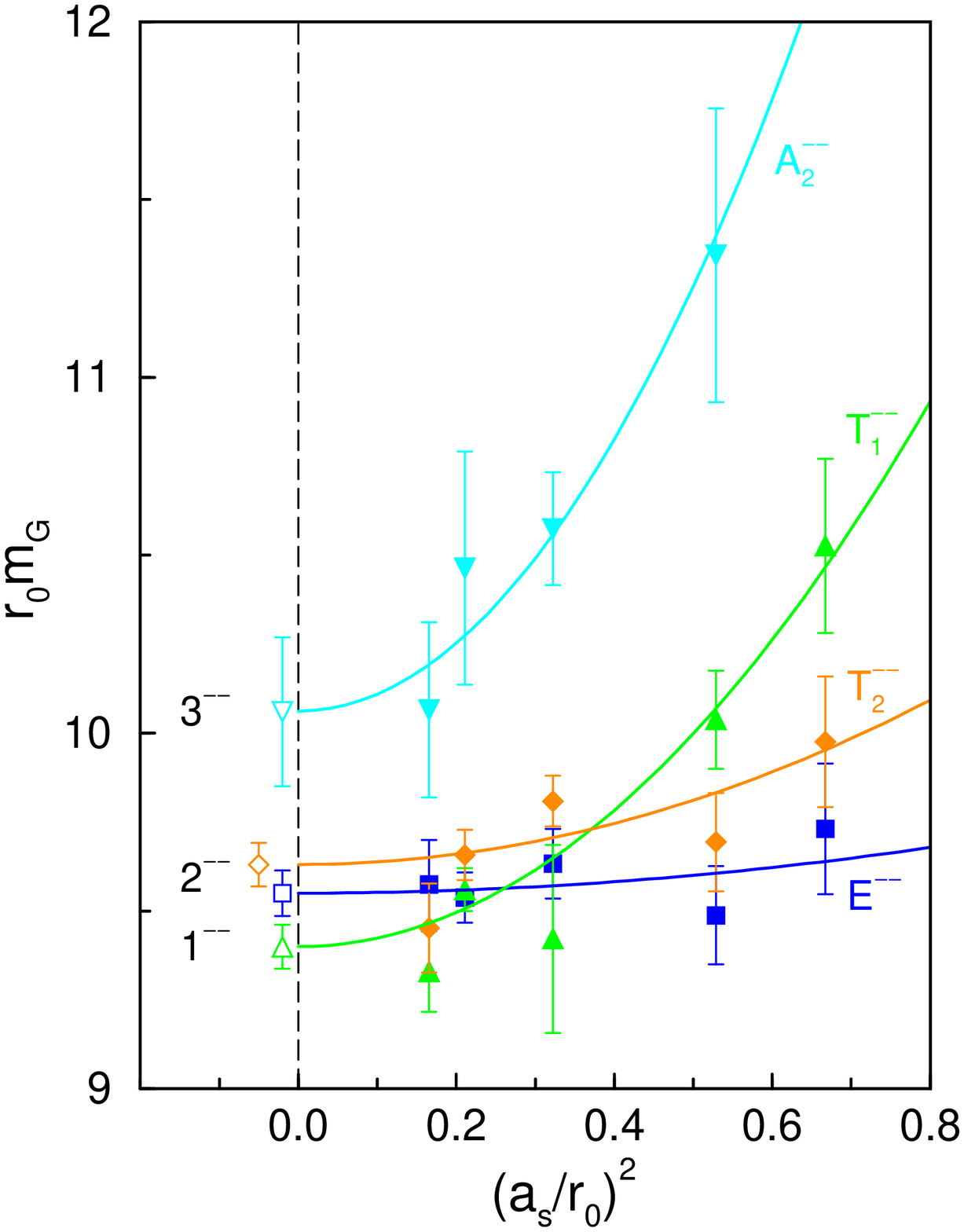}
\end{center}
\caption[figglueMM]{
  Mass estimates (solid symbols) of the $PC=--$ glueballs in terms of $r_0$
  against the lattice spacing $(a_s/r_0)^2$.  The solid curves are best
  fits of $\varphi_0(a_s)$ from Eq.~(\protect\ref{phifour}) to the
  results for each state; the open symbols are the continuum limit
  extrapolations.}
\label{fig:glueMM}
\end{figure}

The glueball mass estimates in terms of $a_t^{-1}$, listed in
Table~\ref{tab:masses}, were combined with the determinations of the
hadronic scale $r_0/a_s$ presented in Table~\ref{tab:runparams}.
The results are shown in Figs.~\ref{fig:gluePP}-\ref{fig:glueMM}.
In these figures, the dimensionless products of $r_0$ and the glueball
mass estimates are shown as functions of $(a_s/r_0)^2$.  The solid
symbols indicate the results from the $\xi=5$ simulations.  The
open symbols appearing to the right of the vertical dashed lines
indicate results from the $\beta=3.0$, $\xi=3$ run, as well as
$\xi=3$ results for the $A_1^{++}$ and $A_1^{\ast ++}$ channels
previously obtained in Ref.~\cite{effglue}.  To remove discretization
errors from our glueball mass estimates, the results for each level
in these figures must be extrapolated to the continuum limit
$a_s/r_0\!\rightarrow\! 0$.  The discretization errors can then
be seen as the deviations of the finite-$a_s$ results from these
limiting values.

From perturbation theory, the leading discretization errors in our
action are expected to be $O(a_t^2,a_s^4,\alpha_s a_s^2)$.  
The agreement of the $A_1^{++}$, $A_1^{\ast ++}$,
$E^{++}$, $T_2^{++}$, and $T_1^{+-}$ glueball masses extracted using
$\xi=3$ and $\xi=5$ (see Figs.~12 and 14 in Ref.~\cite{effglue}) suggests
that the $O(a_t^2)$ errors are negligible.  Some evidence for the smallness
of the $O(\alpha_s a_s^2)$ errors comes from our earlier glueball 
simulations\cite{melbourne} which used the one-loop improved 
L\"uscher-Weisz action\cite{LWaction}.  After tadpole improvement, the
radiative corrections to the couplings in this action were found to
be very small.  For these reasons, we expect that the
$O(a_t^2,\alpha_s a_s^2)$ errors will be negligible compared to the
$O(a_s^4)$ errors. 

In assuming $a_s/a_t=\xi$ (where $\xi$ is the input anisotropy parameter
in the action), we introduce $O(\alpha_s)$ errors in our estimates of
the glueball masses multiplied by $r_0$.  (Note that these errors do not
enter into ratios of the glueball masses.)  Such errors can be avoided
by instead setting $a_s/a_t$ in a more physically motivated fashion, such
as by comparing the spatial and temporal length scales
extracted from appropriate correlation functions.  We can also use 
perturbation theory to adjust the couplings in our action to remove these
errors order by order in $\alpha_s$.  However, we estimated the
errors caused by imposing $a_s/a_t=\xi$ (see below) and found them to
be too small to warrant the additional complexity of another
$a_s/a_t$-setting scheme or the effort required to calculate the
one-loop corrections to the action.  A simpler approach is to
incorporate the $O(\alpha_s)$ anisotropy errors into our continuum limit
extrapolations.  Unfortunately, their dependence on $a_s$ is not well
known and, as we shall see, they are generally smaller than the much more
rapidly varying $O(a_s^4)$ errors. Detecting their effects in a fit to
about five data points for $a_s$ from 0.2 to 0.4 fm is not feasible.  Thus,
we decided to adopt the following approach: to extrapolate assuming $O(a_s^4)$
errors only, then include a systematic uncertainty in our continuum-limit
results from the $O(\alpha_s)$ anisotropy errors.

One way to estimate this uncertainty is to compare measurements of the
static-quark potential extracted from Wilson loops taken along the different
spatial and temporal axes of the lattice\cite{anisopaper}.  The anisotropy
errors can be quantified by defining $a_s/a_t$ in terms of these different
potentials and comparing the result to $\xi$.  If we denote the determination
of the aspect ratio from the potentials by $[a_s/a_t]_V$ and define $Z_\xi$
by the relation $[a_s/a_t]_V=Z_\xi \xi$, then the deviation of $Z_\xi$ from
unity gives us a measure of the fractional error from assuming $a_s/a_t=\xi$. 
The effect of these errors is to modify the multiplicative $r_0/a_t$ factors 
used to convert the simulation results given in terms of $a_t^{-1}$ into 
units of $r_0^{-1}$ suitable for extrapolation. Using the functional dependence of 
$r_0/a_t$ on $\xi$ from a fit to the static-quark potential, we determine 
that $(Z_\xi-1)/2$ gives us an estimate of the fractional uncertainty in our 
continuum limit results from the aspect ratio errors.  Without mean-link 
improvement, $Z_\xi$ can deviate from unity by as much as $30\%$. 
When the action includes mean-link factors, the corrections are found to 
be small, typically a few per cent.  For example, for the 
$\beta=2.4$, $\xi=5$ run, we obtained $Z_\xi=0.987(8)$; for $\beta=3.0$, 
$\xi=3$, an estimate of $Z_\xi=0.99(1)$ was found.  If we assign a 
conservative $2\%$ error from $Z_\xi$, then this amounts to a $1\%$ 
systematic uncertainty in our continuum-limit results from
the anisotropy errors.

Another way to estimate the errors due to the aspect ratio is from the
perturbative calculation of $Z_\xi$ using the anisotropic Wilson action (the
analogous calculation using $S_{I\!I}$ has not yet been done).  Modifying
the results from Ref.~\cite{karsch} to include tadpole improvement
factors and writing the mean-link parameter 
$u_s=1-\alpha_s u_s^{(2)}(\xi)+O(\alpha_s^2)$, one finds
\begin{equation}
Z_\xi = 1 + \alpha_s\, \left\{2\pi\left[c_\tau(\xi)\!-\!c_\sigma(\xi)\right]
+u_s^{(2)}(\xi)\right\} + O(\alpha_s^2),
\end{equation}
where the values for $c_\tau(\xi)$ and $c_\sigma(\xi)$ can be obtained
from Fig.~1 in Ref.~\cite{karsch}.  From this equation, one sees that
when $\xi=3-5$ and $\alpha_s\sim 0.2$, the aspect ratio receives less
than a $2\%$ correction; if the tadpole improvement factor $u_s^{(2)}(\xi)$
is omitted, a large $30\%$ correction is found.  For the improved
action $S_{I\!I}$, we expect that these values should be somewhat smaller.
Again, we can assign a conservative $2\%$ error in $Z_\xi$
to obtain a $1\%$ systematic uncertainty in our continuum estimates.

To summarize, our approach is to extrapolate to the continuum
limit using
\begin{equation}
\varphi_0(a_s) = r_0m_G + c_4\ \frac{a_s^4}{r_0^4},
\label{phifour}
\end{equation}
where $r_0m_G$ and $c_4$ are the best-fit parameters, then add a 
$1\%$ systematic uncertainty from the $O(\alpha_s)$ anisotropy errors.
Eq.~(\ref{phifour}) worked well in all cases except for the $A_1^{++}$
and $A_1^{\ast ++}$ levels. The best-fit curves using Eq.~(\ref{phifour})
are shown in Figs.~\ref{fig:gluePP}-\ref{fig:glueMM}; the extrapolations
of these curves to the continuum limit are indicated in these
figures by the open symbols on the left-hand sides of the vertical
dashed lines.  Note that the extrapolation uncertainties shown in
these plots do not yet include the systematic anisotropy error.

As discussed in Ref.~\cite{effglue}, our results for the  
$A_1^{++}$ and $A_1^{\ast ++}$ levels remain problematical.
These levels have large finite-lattice-spacing errors which do not obey
Eq.~(\ref{phifour}).  There is growing evidence that these large
discretization errors are due to the presence of a critical end point
of a line of phase transitions (not corresponding to any physical
transition found in QCD) in the fundamental-adjoint coupling 
plane\cite{patel,heller,peardon}.  It has been conjectured that
this critical point defines the continuum limit of a $\phi^4$
scalar field theory\cite{heller}.  As one nears this critical
point, the coherence length in the scalar channel becomes large,
which means that the mass gap in this channel becomes small;
all other observables, including glueball masses, appear to be
affected to a much lesser extent.  The effect of this critical
point on the functional form of the discretization errors in the
scalar glueball mass is not well known, so we must proceed somewhat
empirically.  One possibility is that the critical point might
enhance the perturbative $O(\alpha_s)$ errors; another is that
it may induce an $O(a_s^2)$ error which would not otherwise be
present.  We fit several simple functions to the $A_1^{++}$ mass
results and found that the following two functions worked very well:
\begin{eqnarray}
\varphi_1(a_s) &=& r_0m_G + c_4\ \frac{a_s^4}{r_0^4}\nonumber\\
 &&-\left(d_2\ \frac{a_s^2}{r_0^2} + d_4\ \frac{a_s^4}{r_0^4}\right)
     \bigl[\ln(a_s\Lambda)\bigr]^{-1}, \label{phiscalar}\\
\varphi_2(a_s) &=&  r_0m_G + c_2\ \frac{a_s^2}{r_0^2}
   + c_4\ \frac{a_s^4}{r_0^4} + c_6\ \frac{a_s^6}{r_0^6},
\label{phicubic}
\end{eqnarray}
where $r_0m_G$, $c_4$, $c_6$, $d_2$, and $d_4$ are the best-fit parameters.
Eq.~(\ref{phicubic}) is simply a cubic polynomial in $(a_s/r_0)^2$.
Eq.~(\ref{phiscalar}) incorporates the expected leading dependence on the
QCD coupling $\alpha_s(a_s)\sim -1/\ln(a_s\Lambda)$ up to $O(a_s^4)$.  Various
estimates of the QCD scale parameter $\Lambda_{\overline{\rm MS}}$
suggest that $r_0\Lambda_{\overline{\rm MS}}\sim 0.6$.  Hence, we used
$r_0\Lambda=0.5$ and verified that our continuum limit estimates
were insensitive to the choice of $r_0\Lambda$ in the range
from about 0.3 to 0.8.  Both the $d_2$ and $d_4$ terms were needed
to achieve this insensitivity.  

The best-fit curves using Eq.~(\ref{phiscalar}) are shown in 
Fig.~\ref{fig:gluePP}.  The best fit to the $A_1^{++}$ results has
$\chi^2/N_{\rm DF}=0.57$, and for the $A_1^{\ast ++}$, we
find $\chi^2/N_{\rm DF}=0.35$.  The extrapolation of these curves to
the continuum limit yields $r_0m({A_1^{++}})=4.21(7)$ and 
$r_0m({A_1^{\ast ++}})=6.50(44)$.  Using Eq.~(\ref{phicubic}),
we obtain $r_0m({A_1^{++}})=4.30(8)$ with $\chi^2/N_{\rm DF}=0.53$ and
$r_0m({A_1^{\ast ++}})=6.52(54)$ with $\chi^2/N_{\rm DF}=0.35$
Since $\varphi_1(a_s)$ was more closely connected to a perturbative analysis,
we chose the estimates obtained using Eq.~(\ref{phiscalar}) for our
final results, but added the differences between the two extrapolations
as a systematic error.  After including an additional $1\%$ anisotropy
error, we end up with $r_0m(A_1^{++})=4.21(12)$ and 
$r_0m(A_1^{\ast ++})=6.50(45)$.

Note that the $A_1^{++}$ estimate differs slightly from our earlier
estimate of $3.98(15)$ given in Ref.~\cite{effglue}.  Our previous
extrapolation suffered from the absence of a mass measurement at a
lattice spacing smaller than 0.2 fm; the need for such a measurement to
obtain a reliable continuum limit estimate for this level was acknowledged
in Ref.~\cite{effglue}.  The current study includes a new measurement at a
lattice spacing near 0.1 fm; the inclusion of this new measurement is
responsible for the slight difference in the two extrapolations.  Note
that our improved estimate $4.21(12)$ agrees very well with the value
$4.33(5)$ obtained by extrapolating existing Wilson action data for the
scalar glueball mass.  Unfortunately, the mass of the $A_1^{\ast ++}$ is
very poorly determined because a measurement of the $A_1^{\ast ++}$ mass was
not obtained in the $\beta=3.0$, $\xi=3$ simulation.  

Recently, we have demonstrated that the discretization errors in the
scalar glueball mass can be dramatically reduced by simulating with an action
which includes an additional two-plaquette interaction\cite{peardon}.
With such an action, we should be able to substantially improve upon
our determinations of the scalar-channel glueball masses in the near
future.  A study in SU(2) lattice gauge
theory\cite{vancouver} has also shown that lattice-spacing errors in
the scalar glueball can be reduced by using the mean spatial and
temporal links in Landau gauge for the values of the
link variable renormalization parameters $u_s$ and $u_t$, respectively.

\subsection{Spin identification}
\label{sec:spinid}

The last step in our calculation of the glueball spectrum is to
identify the continuum spin content of each level.  This is done
by matching the observed patterns of degeneracies in the levels
from different $O_h^C$ representations to those expected for the
various continuum $J^{PC}$ states.  For example, a $J=0$
state occurs only in the $A_1$ representation of $O$, a $J=2$
state occurs in both the $E$ and $T_2$ representations, and a
state of $J=3$ gets split across the $A_2$, $T_1$, and $T_2$
representations.  The numbers of times that the irreducible 
representations of the octahedral group $O$ occur in the subduced
representations $J\downarrow O$ of the rotation group $SO(3)$
restricted to the subgroup $O$ are listed in Table~\ref{tab:Jsubduced}.
Given the values in this table and either the continuum limit
estimates in Figs.~\ref{fig:gluePP}-\ref{fig:glueMM} or the
results shown in Fig.~\ref{fig:spinanal}, we can then deduce the
continuum spin quantum numbers.

\begin{table}[t]
\caption[tabsix]{
 Number of times each irreducible representation of the
 octahedral group $O$ occurs in the subduced representations
 $J\downarrow O$ of the rotation group $SO(3)$ restricted to
 subgroup $O$.
\label{tab:Jsubduced}}
\begin{center}
\begin{minipage}{2.1 true in}
\begin{tabular}{cccccc}
 $J$ & $A_1$ & $A_2$ & $E$ & $T_1$ & $T_2$ \\
\hline
 0 & 1 & 0 & 0 & 0 & 0 \\
 1 & 0 & 0 & 0 & 1 & 0 \\
 2 & 0 & 0 & 1 & 0 & 1 \\
 3 & 0 & 1 & 0 & 1 & 1 \\
 4 & 1 & 0 & 1 & 1 & 1 \\
 5 & 0 & 0 & 1 & 2 & 1 \\
 6 & 1 & 1 & 1 & 1 & 2 \\
 7 & 0 & 1 & 1 & 2 & 2 \\
 8 & 1 & 0 & 2 & 2 & 2 \\
 9 & 1 & 1 & 1 & 3 & 2 \\
10 & 1 & 1 & 2 & 2 & 3 \\
11 & 0 & 1 & 2 & 3 & 3 \\
12 & 2 & 1 & 2 & 3 & 3
\end{tabular}
\end{minipage}
\end{center}
\end{table}

Consider first the $PC=++$ sector.  The $A_1^{++}$ state is not
degenerate with any other level; hence, it can be identified as
a $J=0$ state.  The $E^{++}$ and $T_2^{++}$ states are degenerate,
implying that they correspond to the five polarizations of a
$J=2$ glueball.  The $A_1^{\ast ++}$ state is seen to have no degenerate
partners, suggesting a $J=0$ excited state.  The $A_2^{++}$ state can 
correspond to $J=3,6,7,9,\dots$.  For all of these $J$ values, there should be 
an accompanying level in the $T_1^{++}$ channel, and such a level is
observed. We conclude that the $A_2^{++}$ and $T_1^{++}$ states correspond
most likely to a $J=3$ state, but less likely $J=6,7,9,\dots$ interpretations
cannot be ruled out. The $J=3$ assignment is also supported by model
predictions, discussed in Sec.~\ref{sec:discuss}. 

In the $PC=-+$ sector, the $A_1^{-+}$ and $A_1^{\ast -+}$ states
are easily identified with $J=0$ states, and the degenerate
$E^{-+}$ and $T_2^{-+}$ states must correspond to a $J=2$ glueball.
The degenerate $E^{\ast -+}$ and $T_2^{\ast -+}$ states most likely
correspond to a $J=2$ state as well, although as noted in 
Sec.~\ref{sec:twoglue}, the proximity of these levels to the two-glueball 
threshold leaves their status somewhat uncertain. Also, we cannot statistically
rule out the possibility that, in combination with two states in the 
$T_1^{-+}$ irreducible representation, they are associated with a $J=5$
glueball.  We cannot rule out an accompanying degenerate level in the
$A_1^{-+}$ channel.  If such a state exists, then these could be the
levels corresponding to a $J=4$ or a $J=8$ glueball.

The $T_1^{+-}$ state must correspond to a $J=1$ glueball, and the
degenerate $A_2^{+-}$, $T_1^{\ast +-}$, and $T_2^{+-}$ levels
correspond to a $J=3$ glueball.  A glueball having $J=2$ is the
most likely interpretation for the degenerate $E^{+-}$ and
$T_2^{\ast +-}$ states.  However, we cannot rule out the possibility
of accompanying levels in the $T_1^{+-}$ and $A_2^{+-}$ channels.
Taking all possibilities into account, the alternate interpretations
are $J=5,7,11$.  The very high lying $A_1^{+-}$ can be interpreted
as a $J=0$ glueball, but $J=4,6,8,\dots$ cannot be excluded.

Finally, consider the $PC=--$ sector.  The most probable scenario is
as follows: the $T_1^{--}$ corresponds to a $J=1$ glueball, the
degenerate $E^{--}$ and $T_2^{--}$ are the five polarizations of a
$J=2$ glueball, and the $A_2^{--}$ is a $J=3$ state.  Of course,
we cannot rule out the presumably higher-lying $J=6,7,9,11,\dots$
interpretations for the $A_2^{--}$.  Another possibility is that the 
$E^{--}$, $T_1^{--}$, and $T_2^{--}$ are degenerate, in which case they could 
correspond to a $J=5$ glueball. Even less likely is that all four levels are 
degenerate. In this case, one could interpret them as a single $J=7$ or
$J=11$ state, or as accidentally degenerate $J=3$ and $J=2$ glueballs.
However, all higher-spin interpretations would require a degenerate 
ground state in either the $T_1^{--}$ or $T_2^{--}$ channel. Our correlator 
fits suggest these degeneracies are missing, making $J=5,6,7 \dots$ 
interpretations unlikely. 

\section{Results and discussion}
\label{sec:discuss}

\newcommand{\bb}{\hspace{1.25ex}}
\begin{table}[t]
\caption[tabseven]{
 Final continuum-limit glueball mass estimates $m_G$.
 When a unique $J$ interpretation for a state cannot
 be made, the other possibilities are indicated in the second column.
 States whose interpretation requires further study are indicated by a dagger. 
 In column 3, the first error is the statistical uncertainty from the
 continuum-limit extrapolation and the second is the estimated uncertainty
 from the anisotropy. In the final column, the first error comes
 from the combined uncertainties in $r_0 m_G$, the second from the
 uncertainty in $r_0^{-1}=410(20)$ MeV.
\label{tab:finalmasses}}
\begin{center}
\begin{minipage}{3.0 true in}
\begin{tabular}{llll}
 $J^{PC}$ & Other $J$ & \hspace{3ex} $r_0 m_G$ & \hspace{2ex} $m_G$ (MeV) \\
\hline
 $0^{++}$  &                & \bb 4.21 (11)(4)            & 1730 (50)\bb(80)  \\
 $2^{++}$  &                & \bb 5.85 (2)\bb(6)          & 2400 (25)\bb(120) \\
 $0^{-+}$  &                & \bb 6.33 (7)\bb(6)          & 2590 (40)\bb(130) \\
 $0^{*++}$ &                & \bb 6.50 (44)(7)$^\dagger$  & 2670 (180)(130)  \\
 $1^{+-}$  &                & \bb 7.18 (4)\bb(7)          & 2940 (30)\bb(140) \\
 $2^{-+}$  &                & \bb 7.55 (3)\bb(8)          & 3100 (30)\bb(150) \\
 $3^{+-}$  &                & \bb 8.66 (4)\bb(9)          & 3550 (40)\bb(170) \\
 $0^{*-+}$ &                & \bb 8.88 (11)(9)            & 3640 (60)\bb(180) \\
 $3^{++}$  & $6,7,9,\dots$  & \bb 8.99 (4)\bb(9)          & 3690 (40)\bb(180) \\
 $1^{--}$  & $3,5,7,\dots$  & \bb 9.40 (6)\bb(9)          & 3850 (50)\bb(190) \\
 $2^{*-+}$ & $4,5,8,\dots$  & \bb 9.50 (4)\bb(9)$^\dagger$& 3890 (40)\bb(190) \\
 $2^{--}$  & $3,5,7,\dots$  & \bb 9.59 (4)\bb(10)         & 3930 (40)\bb(190) \\
 $3^{--}$  & $6,7,9,\dots$  &    10.06 (21)(10)           & 4130 (90)\bb(200) \\
 $2^{+-}$  & $5,7,11,\dots$ &    10.10 (7)\bb(10)         & 4140 (50)\bb(200) \\
 $0^{+-}$  & $4,6,8,\dots$  &    11.57 (12)(12)           & 4740 (70)\bb(230)  
\end{tabular}
\end{minipage}
\end{center}
\end{table}

\begin{figure}[t]
\begin{center}
\leavevmode
\epsfxsize=\columnwidth\epsfbox[29 73 506 611]{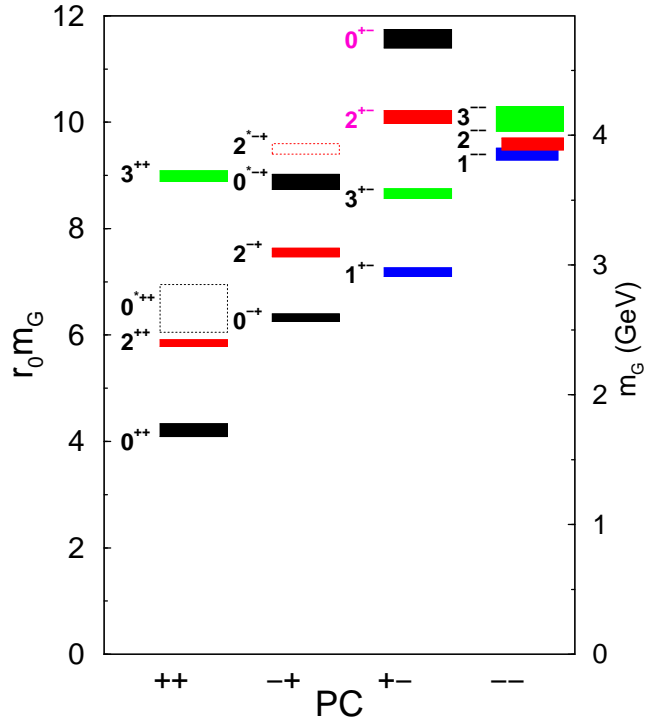}
\end{center}
\caption[figcont]{
 The mass spectrum of glueballs in the pure SU(3) gauge theory.
 The masses are given in terms of the hadronic scale $r_0$ along the
 left vertical axis and in terms of GeV along the right vertical axis
 (assuming $r_0^{-1}=410$ MeV).  The mass uncertainties indicated by the
 vertical extents of the boxes do {\em not} include the uncertainty in
 setting $r_0$.  The locations of states whose interpretation requires further 
 study are indicated by the dashed hollow boxes.}
\label{fig:cont}
\end{figure}

Our final results for the glueball spectrum in terms of $r_0$ are given
in Table~\ref{tab:finalmasses}.  In this table, we assumed the most
likely spin interpretations as described in the previous
section and accordingly combined the continuum
limit extrapolations (shown as open symbols on the left-hand sides
of the vertical dashed lines in Figs.~\ref{fig:gluePP}-\ref{fig:glueMM}),
then added the $1\%$ anisotropy error, to obtain final estimates for
the glueball masses in terms of $r_0$.  The combinations
used to obtain these estimates are also indicated in
Figs.~\ref{fig:gluePP}-\ref{fig:glueMM} by the $J^{PC}$ labels near the
left vertical axes.  Wherever applicable, we have also indicated
in Table~\ref{tab:finalmasses} any alternative spin interpretations
which cannot be ruled out.  These final estimates are also shown in
Fig.~\ref{fig:cont}.  The $0^{\ast ++}$ and $2^{\ast -+}$ states are
shown as dashed hollow boxes to indicate that their interpretations
as glueballs are tentative.  Our concern about the $0^{\ast ++}$ state
stems from its non-negligible finite volume effects; for the $2^{\ast -+}$
state, its nearness to the two-glueball threshold in our simulations is 
worrisome.  Note that our estimates $r_0m(0^{++})=4.21(12)$ and
$r_0m(2^{++})=5.85(6)$ agree very well with $4.33(5)$ and $6.0(1)$,
respectively, obtained by extrapolating the Wilson action simulation
results from Refs.~\cite{ukqcd,teper,gf11,forcrand} to the continuum limit.
Several glueball mass ratios are presented in Table~\ref{tab:massratios}.
We can determine these ratios very accurately since, as noted earlier,
they are not contaminated by anisotropy errors.  The uncertainties given
in Table~\ref{tab:massratios} are calculated using the empirical fact that
correlations between different symmetry channels were found to be negligible.
Note that the pseudoscalar glueball is clearly resolved (at the $7\sigma$
level) to be heavier than the tensor. 

\begin{table}
\caption[tabeight]{
  Glueball mass ratios. 
\label{tab:massratios}}
\begin{center}
\begin{minipage}{2.0 true in}
\begin{tabular}{ll}
$m(2^{++})/m(0^{++})$     &     $1.39(4)$ \\
$m(0^{-+})/m(0^{++})$     &     $1.50(4)$ \\
$m(0^{\ast++})/m(0^{++})$  &     $1.54(11)$ \\
$m(1^{+-})/m(0^{++})$     &     $1.70(5)$ \\
$m(2^{-+})/m(0^{++})$     &     $1.79(5)$ \\
$m(3^{+-})/m(0^{++})$     &     $2.06(6)$\\
$m(0^{\ast-+})$/$m(0^{++})$  &  $2.11(6)$ \\[2mm]
$m(0^{-+})$ /$m(2^{++})$     &  $1.081(12)$
\end{tabular}
\end{minipage}
\end{center}
\end{table}

All of the glueball states shown in Fig.~\ref{fig:cont} are stable
against decay to lighter glueballs.  In the $PC=++$ sector, the threshold for
decay into two identical $0^{++}$ glueballs having zero total momentum
is at twice the mass of the scalar glueball.  Although this lies below
the mass of the $3^{++}$ glueball, Bose symmetrization prohibits odd $L$
partial waves, where $L$ is the relative orbital angular momentum,
so that the $3^{++}$ glueball cannot decay into two
identical scalar glueballs.  In the $PC=-+$ sector, the lowest-lying
two-glueball state consists of the $0^{++}$ and $2^{++}$ glueballs in
a relative $P$-wave; all of our glueballs in this sector have masses below
the sum of the scalar and tensor glueball masses.  States of total
zero momentum and comprised of the $0^{++}$ and $1^{+-}$ glueballs with
relative orbital angular momentum $L$ are the lowest-lying two-glueball
states in the $PC=+-$ sector when $L$ is even and in the $PC=--$ sector
when $L$ is odd.  Only the $0^{+-}$ glueball has sufficient mass to decay
into two such glueballs; however, this decay is forbidden because $L=1$
is required to make a state of zero total angular momentum.

To convert our glueball masses into physical units, the value of the
hadronic scale $r_0$ must be specified.  We used $r_0^{-1}=410(20)$ MeV
from Ref.~\cite{effglue} to obtain the scale shown on the right-hand
vertical axis of Fig.~\ref{fig:cont}.  This estimate was obtained by
combining Wilson action calculations of $a/r_0$ with values of the lattice
spacing $a$ determined using quenched simulation results of various physical
quantities, such as the masses of the $\rho$ and $\phi$ mesons, the decay
constant $f_\pi$, and the $1P-1S$ splittings in charmonium and bottomonium.
Note that the errors shown in Fig.~\ref{fig:cont} do not include the
uncertainty in $r_0^{-1}$.  For the lowest-lying glueballs, we obtain
$m(0^{++})=1730(50)(80)$ MeV and $m(2^{++})=2400(25)(120)$ MeV, where the
first error comes from the uncertainty in $r_0m_G$ and the second error
comes from the uncertainty in $r_0^{-1}$.  A great deal of care should be
taken in making direct comparisons with experiment since these values
neglect the effects of light quarks and mixings with nearby conventional
mesons.  It is this mixing which has made the search for an incontrovertible
experimental signal so difficult.  A glueball having exotic $J^{PC}$ 
will not mix with conventional hadrons and would be ideal for establishing
the existence of glueballs.  Unfortunately, our results indicate that the
lightest such state, the $2^{+-}$ glueball, has a mass greater than 4 GeV. 

Kuti has recently pointed out\cite{kuti} that the glueball spectrum
shown in Fig.~\ref{fig:cont} can be qualitatively understood
in terms of the interpolating operators of minimal dimension which can
create glueball states.  With the expectation that higher dimensional
operators create higher mass states, the authors in Ref.~\cite{jaffe},
following an approach suggested in Refs.~\cite{fritzsch,bjorken},
constructed all operators of dimension $4$, $5$, and $6$ capable of
creating glueballs from the QCD vacuum.  Such operators are
gauge-invariant combinations of the chromoelectric and chromomagnetic
fields; operators equivalent to a total derivative or related to a
conserved current are excluded.  The lowest dimensional operators capable
of creating glueballs are of dimension four and have the form
${\rm Tr} F_{\mu\nu} F_{\alpha\beta}$, where $F_{\mu\nu}$ is the gauge field
strength tensor; these operators create glueballs with $J^{PC} =
0^{++},2^{++},0^{-+}$ and $2^{-+}$.   The dimension five operators
of the form ${\rm Tr} F_{\mu\nu}D_\delta F_{\alpha\beta}$, where $D_\mu$
is the covariant derivative, produce only two new glueball states
having $J^{PC}=1^{++}$ and $3^{++}$.  At dimension six, operators of
the form ${\rm Tr} F_{\mu\nu} F_{\alpha\beta} F_{\delta\sigma}$
produce $J^{PC}=0^{\pm +},1^{\pm\pm},2^{\pm\pm}$, and $3^{\pm -}$;
operators of the form ${\rm Tr} F_{\mu\nu} \{D_\alpha,D_\beta\}
F_{\delta\sigma}$ produce $J^{PC}=1^{-+},3^{-+}$, and $4^{\pm +}$.
Of course, this ordering should not be taken too quantitatively, but
we find that the method provides a reasonably satisfactory explanation
of the observed spectrum, especially given the simplicity of the approach.
Of the lightest six states we resolve, four have the quantum numbers
expected from the dimension-four interpolating operators.
The method also explains the absence of any low-lying $0^{\pm -}$ and
$1^{-+}$ glueballs.

The spectrum of Fig.~\ref{fig:cont} can also be reasonably well explained
in terms of a simple constituent gluon model in which the fundamental
gluon field is replaced by the Hartree modes of a constituent field
with residual perturbative interactions; the Hartree modes are taken
to be the modes of a free gluon inside a spherical cavity with confining
boundary conditions.  Such a (bag) model has been recently revisited in
Ref.~\cite{kuti}.  Using values for $\alpha_s$ and the bag pressure
appropriate for heavy-quark spectroscopy and the static-quark potential,
remarkable agreement with the observed levels of Fig.~\ref{fig:cont} was
found.

\section{Conclusion}
\label{sec:conclude}

In this paper, we used numerical simulations of gluons on spatially-coarse,
temporally-fine lattices to significantly improve our knowledge of the
glueball spectrum in SU(3) Yang-Mills theory.  This is an important
step towards understanding glueballs in the real world.  Six simulations
for spatial grid separations ranging from 0.1 to 0.4 fm were performed
on DEC Alpha and Sun Ultrasparc workstations.  Care was taken to differentiate
single glueball states from unwanted two-glueball and torelon-pair states.
An additional small-volume simulation assisted in the identification of the
single glueball states and demonstrated the smallness of systematic errors
from finite volume.  The simulation results were extrapolated to the
continuum limit and the continuum spin quantum numbers were identified.
The end result, shown in Fig.~\ref{fig:cont}, was a nearly complete survey
of the glueball spectrum in the pure glue theory below 4 GeV.  A total of
thirteen glueballs were found, and two other tentative candidates were also
located.  

In the future, we plan to improve our determinations of the scalar-channel
glueballs by simulating with an action that includes an additional
two-plaquette interaction.  We also intend to extend our anisotropic
lattice technology to include quarks.  With the help of femto-universe
techniques, we hope ultimately to investigate the properties of glueballs
in reality.

\section*{Acknowledgments}
We would like to thank Peter Lepage, Julius Kuti, Mike Teper, Ron Horgan and 
Chris Michael for helpful discussions.  We are grateful to W.~Korsch 
(University of Kentucky) for access to computing resources in the early stages
of the project.  This work was supported by the U.S.~DOE, Grant
No.~DE-FG03-97ER40546.

%%%%%%%%%%%%%%%%%%%%%%%%%%%%%%%%%%%%%%%%%%%%%%%%%%%%%%%%%%%%%%%%%%%%%
%%
%%                             REFERENCES
%%
%%%%%%%%%%%%%%%%%%%%%%%%%%%%%%%%%%%%%%%%%%%%%%%%%%%%%%%%%%%%%%%%%%%%%

%%%%%%%%%%%%%%%%%%%%%%%%%%%%%%%%%%%%%%%%%%%%%%%%%%%%%%%%%%%%%%%%%%%%%
\end{document}